\newcommand{\be}{\begin{equation}}
\newcommand{\ee}{\end{equation}}

\def\spose#1{\hbox to 0pt{#1\hss}}
\def\ltapprox{\mathrel{\spose{\lower 3pt\hbox{$\mathchar"218$}}
 \raise 2.0pt\hbox{$\mathchar"13C$}}}
\def\gtapprox{\mathrel{\spose{\lower 3pt\hbox{$\mathchar"218$}}
 \raise 2.0pt\hbox{$\mathchar"13E$}}}
\def\reff#1{(\ref{#1})}
\newcommand{\ba}{\begin{eqnarray}}
\newcommand{\ea}{\end{eqnarray}}
\documentstyle[12pt,epsf]{article}
\begin{document}
\begin{titlepage} 
\vspace*{1.0cm}
\begin{center} {\LARGE \bf  
Infrared Behavior \\
of the Gluon Propagator  \\
in Lattice Landau Gauge\\} \vspace*{0.5cm}
{\bf Attilio Cucchieri$^{(a)}$}\\ \vspace*{0.7cm}
Gruppo APE -- Dipartimento di Fisica \\
Universit\`a di Roma ``Tor Vergata'' \\
Via della Ricerca Scientifica 1, I-00133 Roma, ITALY \\ \vspace*{0.5cm}
May 13, 1997 \\ \vspace*{0.9cm}
{\bf   Abstract   \\ } \end{center} \indent

We evaluate numerically the four-momentum-space gluon propagator
in lattice Landau gauge, for pure $SU(2)$ lattice gauge theory
with periodic boundary conditions.
Simulations are done
in the strong-coupling regime,
namely at $\beta = 0\mbox{,}\, 0.8\mbox{,}\, 1.2$ and
$1.6$, and for lattice sizes up to $30^4\mbox{,}\,
16^4\mbox{,}\, 20^4$ and $24^4$ respectively.
In the limit of large lattice volume,
we observe in all cases a
gluon propagator decreasing as
the momentum goes to zero.
This counter-intuitive result has a straightforward
interpretation as resulting from the proximity of the
so-called first Gribov horizon in infrared directions.

\vfill \begin{flushleft} 
PACS numbers: 11.15.Ha, 12.38.Aw \\
keywords: lattice gauge theory, Landau gauge,
gluon propagator \vspace*{0.7cm} \\
\noindent{\rule[-.3cm]{5cm}{.02cm}} \\
\vspace*{0.2cm} \hspace*{0.5cm} ${}^{a)}$
E-mail address: {\tt cucchieri@roma2.infn.it}. \end{flushleft} 
\end{titlepage}

\noindent
{\bf 1.\hskip0.3cm}
A non-perturbative investigation of QCD in the infrared limit
is necessary in order to get an understanding of crucial
phenomena such as color confinement and hadronization.
In particular, the infrared behavior
of the gluon propagator can be directly related \cite{West}
to the behavior of the Wilson loop at large separations,
and to the existence of an area law.
Of course in developing non-perturbative techniques for QCD,
one has to deal with the redundant gauge degrees
of freedom. The standard gauge-fixing technique
is based on the
assumption that a gauge-fixing condition can be found which
uniquely determines a gauge field on each gauge orbit.
The correctness of this procedure was questioned by Gribov
\cite{Gr}. He showed that the Coulomb and the Landau
gauges do not fix the gauge fields uniquely,
namely there exist gauge-related field configurations
({\em Gribov copies}) which all satisfy the gauge condition.
Gribov copies affect also lattice numerical simulations
\cite{MPR}. In fact, the lattice Landau (or Coulomb) gauge
is imposed by minimizing a functional which usually
displays several relative minima ({\em lattice Gribov copies}).

In order to get rid of the problem of spurious gauge copies, Gribov
proposed \cite{Gr} the use of additional gauge conditions. In
particular he restricted the physical configuration
space to the region $\Omega$ of transverse configurations
({\em i.e.}\ $\partial \cdot A \,=\,0\,$), for which
the Faddeev-Popov operator
$\, {\cal M}\left[A\right]\,\equiv\,- \nabla\cdot\,D\left[A\right]\,$
is nonnegative. This region is delimited by the so-called {\em first
Gribov horizon}, defined as the set of configurations
for which
the smallest, non-trivial eigenvalue\footnote{~The
Faddeev-Popov operator has a
trivial null eigenvalue, corresponding to a constant eigenvector.}
of the Faddeev-Popov operator is zero.
We now know that $\Omega$ is {\em not} free of Gribov copies
and that the physical configuration
space has to be identified with
the so-called {\em fundamental modular region}
\cite{STSF,Z2}. Nevertheless, the region $\Omega$ is of interest in
numerical simulations. In fact, configurations which satisfy
the usual lattice Landau gauge condition belong to this region.

The restriction of the path integral, which defines the
partition function, to the region $\Omega$ implies a 
{\em rigorous} inequality \cite{DZ,Z1} for the Fourier
components of the gluon field. From this
inequality, which is a consequence only of the
positiveness of the Faddeev-Popov operator, it follows \cite{DZ,Z1}
that the region $\Omega$ is bounded by a certain ellipsoid $\Theta$.
This bound implies
proximity of the first Gribov horizon in infrared
directions, and consequent suppression of the low-momentum
components of the gauge field, a result
already noted by Gribov in reference \cite{Gr}.
This bound also causes a strong 
suppression of the gluon propagator in the infrared
limit. More precisely,
Zwanziger proved \cite{Z1} that, in the infinite-volume limit,
the gluon propagator is less singular
than $k^{-2}$ in the infrared limit
and that, very likely, it does vanish as rapidly as $k^{2}$.
A gluon propagator that vanishes as $k^{2}$ in the infrared
limit was also found (under certain hypotheses) by
Gribov \cite{Gr}. In particular, he obtained
$ D(k) = k^{2} / ( k^{4} \,+\, \gamma ) $.
The momentum scale $\gamma$ appears when
the restriction of the physical configuration space to the
region $\Omega$ is imposed. This propagator agrees with the zeroth-order
perturbative prediction $k^{- 2}$ at large momenta, but
gives a null propagator at $k = 0$.
A propagator which is a generalization of the one obtained by Gribov
has also been introduced in reference \cite{Stingl_orig} as an Ansatz for a
non-perturbative solution of the Schwinger-Dyson equation.
Let us notice that these results 
are in complete contradiction with the $k^{- 4}$
singularity obtained for the gluon propagator
when the Schwinger-Dyson equation is {\em approximately solved}
in the infrared limit \cite{BPenn}.

The infrared behavior of the gluon propagator in lattice Landau gauge
has been the subject of relatively few numerical
studies \cite{MO2,Gup,Parrinello,Nakamura,gut}.
In some cases \cite{Parrinello,gut} results
seem to indicate that the gluon propagator is
finite at zero momentum and in the infinite-volume limit.
This is in agreement with Zwanziger's prediction
of a gluon propagator less singular than $k^{-2}$.
Recently, a gluon propagator evaluated avoiding Gribov copies
\cite{Nakamura} has been successfully fitted by a Gribov-like
formula \cite{St}.
Finally, in references \cite{Athesis,C_GN}, it was checked
that the influence of Gribov copies on the
gluon propagator ({\em Gribov noise}) is
of the order of magnitude of the numerical accuracy. In
particular, this seems to be the case even
at small values of the coupling $\beta$, namely in the strong-coupling
regime, where the number of Gribov copies is higher and Gribov noise, if
present, should be larger and more easily detectable.

In references \cite{Athesis,C_GN}
it was also observed, at $\beta = 0$, that the
gluon propagator
{\em decreases} as the momentum goes to zero.
A similar behavior seemed to be present also at
higher values of $\beta$, and in particular
at $\beta = 1.6$ and for lattice volume $V = 24^{4}$.
In the last case, however, the behavior is
not as clear as that at $\beta = 0$,
due to the limited statistics (only $7$
configurations). In the present work we
extend the analysis done in references \cite{Athesis,C_GN},
and we would like to obtain a conclusive result for the
infrared behavior of the gluon propagator
in the strong-coupling regime. 
Let us notice that Zwanziger's predictions \cite{Z2,Z1}
for the gluon propagator are
{\em $\beta$-independent}: in fact, they are
derived only from the positiveness of the
Faddeev-Popov operator when the
lattice Landau gauge is imposed.
Thus, results in the strong-coupling regime are
a valid test for those predictions.
Of course, once the behavior of the gluon
propagator is clarified in this
regime, we should try to extend the results
to larger
values of $\beta$, possibly up to the scaling region.

\vskip1cm
\noindent
{\bf 2.\hskip0.3cm}
We consider a standard Wilson action for
$SU(2)$ lattice gauge theory in $4$ dimensions
with periodic boundary conditions.
For notation and details about numerical 
simulations
we refer to \cite{Athesis,C_GN}.
The only difference will be that here we consider
lattices with different sizes $N_{\mu}$ in the different directions.
Let us recall that the
gluon propagator, which will be evaluated in lattice
Landau gauge \cite{MO2,W2}, is defined, in momentum space,
as\footnote{~Here $k$ has
components $ k_{\mu}\,N_{\mu}  \equiv
0\mbox{,}\,1\mbox{,}\,\ldots \mbox{,}\, N_{\mu} - 1 $.
As in reference \cite{C_GN}, we evaluate the gluon
propagator by considering only values of
$k$ with three of the four components equal to zero, namely
$k = (0\mbox{,}\, 0\mbox{,}\, 0\mbox{,}\, k_{4})$.
\protect\label{foot} }
\ba
D(0)& \equiv& \frac{1}{12 V} \sum_{\mu\mbox{,}\,a}\,\langle\,
  \left[\,\sum_{x}\,A_{\mu}^{a}(x)\,\right]^{2} \,\rangle 
\label{eq:D0def} \\
D(k) & \equiv & \frac{1}{9 V} \sum_{\mu\mbox{,}\,a}\,\langle\,
\left\{\,\left[\,\sum_{x}\,A_{\mu}^{a}(x)\,
\cos{( 2 \pi  k \cdot x )}\,\right]^{2} \right. \nonumber \\
& & \left. \qquad \quad \;
 \,+\,
   \left[\,\sum_{x}\,A_{\mu}^{a}(x)\,
\sin{( 2 \pi  k \cdot x )}\,\right]^{2}
\, \right\} \,\rangle
\label{eq:Dkdef}
\;\mbox{.}
\ea

In Table \ref{Table:thermalization} we report,
for each pair $(\beta\mbox{,}\,N)$, the parameters used
for the simulations. Only for the
lattice volume $V = 8^{4}$, at $\beta = 0.8$
and $1.6$, and for $V = 24^{4}$, at $\beta = 1.6$,
we used the data reported in reference
\cite[average ``fm'']{C_GN}.
Computations were performed on
the IBM SP2 at the Cornell Theory Center, on
several IBM RS-6000/250--340 workstations
at New York University, on a IBM RS-6000/550E
workstation at the University
of Rome ``La Sapienza'' \footnote{~I thank
S.Petrarca and B.Taglienti for kindly offering
me the access to this machine.}, and on an ALPHAstation 255 at 
the ZiF-Center in Bielefeld.

In Figure \ref{fig:gluon}
we plot the data
for the gluon propagator as a function
of the square of the lattice momentum\footnote{~As
said in footnote \ref{foot}, we take
$k = (0\mbox{,}\, 0\mbox{,}\, 0\mbox{,}\, k_{4})$.}
\be
p^{2}(k)
\equiv 4 \sum_{\mu} \sin^{2}{\left( \pi \,k_{\mu} \right)}
\,=\, 4\,\sin^{2}{\left( \pi \,k_{4}\right)}
\label{eq:p2}
\ee
for different lattice sizes, at $\beta = 0\mbox{,}\,
0.8 \mbox{,}\, 1.2$ and $1.6$
respectively.\footnote{~For clarity, at $\beta = 0$ we plot only
the data for the three largest volumes; results are similar
for smaller volumes. For the same reason, at $\beta = 1.6$,
we do not show the data for the lattice volume
$V = 12^{3}$x$24$ which are
very similar to the data for $V = 16^{4}$.}
Finally, in Table \ref{Table:D0_00}, we report
the data for $D(0)$ at $\beta = 0$
as a function of the lattice volume.

Data at $\beta = 0$ confirm previous results \cite{Athesis,C_GN}:
the propagator is decreasing (more or less monotonically)
as $k$ decreases.
At $\beta > 0$ the propagator is also decreasing
in the infrared limit, at least for values of $p^{2}$
smaller than a turn-over\footnote{~A drastic
change in the behavior of $D(k)$, around
a turn-over momentum $p_{to}$, has also been observed
in reference \cite{gut} for values of $\beta$
in the scaling region.} value $p^{2}_{to}$. 
Clearly $p^{2}_{to}$ is $\beta$-dependent. Using,
for each $\beta$, the results for the largest
lattice volume available we obtain:
$p^{2}_{to} \approx 1.4$ at $\beta = 0.8$,
$p^{2}_{to} \approx 1$ at $\beta = 1.2$,
and $p^{2}_{to} \approx 0.6$ at $\beta = 1.6$.

As $\beta$ increases the behavior of the
gluon propagator depends strongly on the lattice
volume. This is evident at $\beta = 1.6$.
In fact, for the smallest lattice volume $V = 8^{4}$, we obtain a 
gluon propagator increasing
(monotonically) as $k$ decreases.
On the contrary, for
$V = 24^{4}$, the gluon propagator clearly decreases
as $p^{2}$ goes to zero. 

Let us notice
that, at high momenta, there are very small finite-size
effects, at all values of $\beta$.
The situation
is completely different in the small-momenta sector, as
already stressed for the case $\beta = 1.6$, and as can be
observed in Figure \ref{fig:gluon} also for
$\beta = 0.8$ and $1.2$.
The value $D(0)$
of the gluon propagator at zero momentum is also very
volume-dependent\footnote{~Only
at $\beta = 1.2$ the value of $D(0)$ is
essentially constant,
within error bars, for the three
lattice volumes considered.}
(see Table \ref{Table:D0_00}
for $\beta = 0$, and Figure \ref{fig:gluon} for the
other values of the coupling). 
In particular, at $\beta = 0$, as the volume increases,
we obtain an initial monotonic decrease of the value of $D(0)$;
for $V = 18^{3}$x$36$ this value unexpectedly jumps up,
and then starts again to decrease monotonically.\footnote{~Notice
that $D(0)$ gets the same value on different lattices
(such as $16^{3}$x$32$ and $20^{4}$, or
$20^{3}$x$40$ and $24^{4}$) with ``similar'' lattice
volume (respectively $131,072$ and $160,000$, or
$320,000$ and $331,776$).}
Finally, for $\beta = 0.8$ and $1.6$,
the value of the zero-momentum gluon propagator
decreases (more or less monotonically)
as $V$ increases. 
These results suggest a finite
value for $D(0)$ in the infinite-volume limit, in
agreement with
references \cite{Parrinello,gut}.
From our data it is not clear if this value would be
zero or a strictly positive constant. Therefore, the
possibility of a zero value for $D(0)$ in the infinite-volume
limit is not ruled out.

We think that our data for the gluon propagator
are very interesting.
The prediction \cite{Gr,Z2,Z1} of a propagator decreasing
as the momentum goes to zero is clearly verified
numerically for several values 
$\beta$, even if only in the strong-coupling
regime. Moreover, it appears that the
lattice size at which this behavior for the gluon
propagator starts to be observed increases with the
coupling.\footnote{~This was expected
since Zwanziger's predictions \cite{Z2,Z1}
for the gluon propagator
are valid only in the infinite-volume limit.}
Of course we should extend our
simulations to higher values of $\beta$
and to larger volumes.
This is, at the moment, beyond the limits
of our computational resources.

\vskip8mm
\noindent
I am indebted to D.Zwanziger for suggesting
this work to me. I also would like to thank him,
G.Dell'Antonio, T.Mendes, V.K.Mitrjushkin, S.Petrarca,
M.Schaden and B.Taglienti for valuable discussions
and suggestions.
I thank for the hospitality
the Physics Department
of the University of Bielefeld and the Center
for Interdisciplinary Research (ZiF),
where part of this work was done.
I also thank for the hospitality
the Physics Department
of the University of Rome ``Tor Vergata'',
and in particular R.Petronzio and the APE
group, where this work was finished.

\clearpage
 

%
\begin{table}
%
%
\protect\footnotesize
\begin{center}
\begin{tabular}{|| c | c | c | c | c | c ||}
\hline
\hline
$ \beta $ & $ V $ & config.\ & therm.\ sweeps & sweeps & $p$ \\ \hline
$ 0.0 $ & $ 12^{4} $ & $  100 $ & $ 4 $ & $ 2 $ & $ 0.945 $ \\ \hline
$ 0.0 $ & $ 16^{4} $ & $  100 $ & $ 4 $ & $ 2 $ & $ 0.96  $ \\ \hline
$ 0.0 $ & $ 16^{3}$x$32 $ & $ 150 $ & $ 2 $ & $ 2 $ & $ 0.97 $ \\ \hline
$ 0.0 $ & $ 20^{4}$ & $ 100 $ & $ 4 $ & $ 2 $ & $ 0.97  $ \\ \hline
$ 0.0 $ & $ 18^{3}$x$36 $ & $ 150 $ & $ 2 $ & $ 2 $ & $ 0.9675 $ \\ \hline
$ 0.0 $ & $ 20^{3}$x$40 $ & $ 150 $ & $ 2 $ & $ 2 $ & $ 0.9725 $ \\ \hline
$ 0.0 $ & $ 24^{4} $ & $ 220 $ & $ 4 $ & $ 2 $ & $ 0.975 $ \\ \hline
$ 0.0 $ & $ 28^{4} $ & $ 150 $ & $ 2 $ & $ 2 $ & $ 0.97 $ \\ \hline
$ 0.0 $ & $ 24^{3}$x$48 $ & $ 90 $ & $ 2 $ & $ 2 $ & $ 0.97 $ \\ \hline
$ 0.0 $ & $ 30^{4} $ & $ 70 $ & $ 2 $ & $ 2 $ & $ 0.975 $ \\ \hline \hline
$ 0.8 $ & $  8^{4} $ & $ 200 $ & $ 1000 $ & $ 100 $ & $ 0.875 $ \\ \hline
$ 0.8 $ & $ 12^{4} $ & $ 200 $ & $ 1375 $ & $ 125 $ & $ 0.9   $ \\ \hline
$ 0.8 $ & $ 16^{4} $ & $  96 $ & $ 1100 $ & $ 100 $ & $ 0.94  $ \\ \hline \hline
$ 1.2 $ & $ 12^{4} $ & $ 150 $ & $ 1000 $ & $ 100 $ & $ 0.92  $ \\ \hline
$ 1.2 $ & $ 16^{4} $ & $ 100 $ & $ 1000 $ & $ 100 $ & $ 0.93  $ \\ \hline
$ 1.2 $ & $ 20^{4} $ & $ 85 $ & $ 1000 $ & $ 100 $ & $ 0.94  $ \\ \hline \hline
$ 1.6 $ & $ 8^{4} $ & $ 200 $ & $ 1100 $ & $ 100 $ & $ 0.865 $ \\ \hline
$ 1.6 $ & $ 12^{3}$x$24 $ & $ 200 $ & $ 1650 $ & $ 150 $ & $ 0.91 $ \\ \hline
$ 1.6 $ & $ 16^{4} $ & $  86 $ & $ 1650 $ & $ 150 $ & $ 0.92  $ \\ \hline
$ 1.6 $ & $ 24^{4} $ & $  34 $ & $ 1650 $ & $ 150 $ & $ 0.92 $ \\ \hline \hline
\end{tabular}
\end{center}
\caption{~The pairs $(\beta\mbox{,} V)$ used
        for the simulations, the number of
         configurations, the number of sweeps used for thermalization,
         the number of sweeps between two consecutive
         configurations used for collecting our data, and
         the parameter $p$ used by the
         stochastic overrelaxation algorithm.}
\label{Table:thermalization}
\vspace*{0.3cm}
\end{table}
\begin{table}
%
\hspace*{-1.0cm}
\protect\footnotesize
\begin{center}
\begin{tabular}{|| c | c ||}
\hline
\hline
$ V $ & $ D(0) $ \\
\hline
$ 12^{4} $ & $ 0.154 (0.007) $ \\ \hline 
$ 16^{4} $ & $ 0.149 (0.006) $ \\ \hline 
$ 16^{3}$x$32 $ & $ 0.143 (0.005) $ \\ \hline 
$ 20^{4} $ & $ 0.142 (0.005) $ \\ \hline 
$ 18^{3}$x$36 $ & $ 0.160 (0.005) $ \\ \hline 
$ 20^{3}$x$40 $ & $ 0.153 (0.006) $ \\ \hline 
$ 24^{4} $ & $ 0.153 (0.004) $ \\ \hline 
$ 28^{4} $ & $ 0.146 (0.005) $ \\ \hline 
$ 24^{3}$x$48 $ & $ 0.144 (0.007) $ \\ \hline 
$ 30^{4} $ & $ 0.136 (0.007) $ \\ \hline
\hline
\end{tabular}
\end{center}
\caption{~The zero four-momentum gluon propagator
          $D(0)$ [see eq.\ \protect\reff{eq:D0def}]
          as a function of the lattice
          volume $V$ at $\beta = 0$.
          Notice that $16^{3}$x$32 \,\approx\,
          20^{4}$ and that $20^{3}$x$40 \, \approx\,
          24^{4}$.}
\label{Table:D0_00}
\vspace*{0.3cm}
\end{table}

\clearpage


%
\begin{figure}[p]
\begin{center}
\vspace*{0cm} \hspace*{-0cm}
\epsfxsize=0.48\textwidth
\leavevmode\epsffile{plot_0.ps}
\hspace{0.3cm}
\epsfxsize=0.48\textwidth
\epsffile{plot_8.ps}  \\
\vspace*{0.5cm}
\epsfxsize=0.48\textwidth
\leavevmode\epsffile{plot_12.ps}
\hspace{0.4cm}
\epsfxsize=0.48\textwidth
\epsffile{plot_16.ps}  \\
\end{center}
\vspace{-0.5cm}
\caption{~Plot of the gluon propagator $D(k)$
[see eqs.\ \protect\reff{eq:D0def} and \protect\reff{eq:Dkdef}]
as a function of the square of the
lattice momentum $p^{2}(k)$ [see eq.\ \protect\reff{eq:p2}]
for lattice volumes:
     ({\bf a}) $V = 28^{4}$ ($\Box$), $V = 24^{3}$x$48$ ($\Diamond$)
         and $V = 30^{4}$ ($\ast$), at $\beta = 0$;
     ({\bf b}) $V = 8^{4}$ ($\Box$), $V = 12^{4}$ ($\Diamond$)
         and $V = 16^{4}$ ($\ast$), at $\beta = 0.8$;
     ({\bf c}) $V = 12^{4}$ ($\Box$), $V = 16^{4}$ ($\Diamond$)
         and $V = 20^{4}$ ($\ast$), at $\beta = 1.2$;
     ({\bf d}) $V = 8^{4}$ ($\Box$), $V = 16^{4}$ ($\Diamond$)
         and $V = 24^{4}$ ($\ast$), at $\beta = 1.6$.
In all our runs we set $k = (0\mbox{,}\, 0\mbox{,}\, 0\mbox{,}\,
k_{4})$.
Since we use periodic boundary conditions,
only data for $k_{4} \leq 1/2$ are reported here.
Error bars are one standard deviation.
}
\label{fig:gluon}
\end{figure}

\end{document}